# Photonic-electronic spiking neuron with multi-modal and multi-wavelength excitatory and inhibitory operation for high-speed neuromorphic sensing and computing

Weikang Zhang[1], Matěj Hejda[1], Qusay Raghib Ali Al-Taai[2], Dafydd Owen-Newns[1], Bruno Romeira[3], José M. L. Figueiredo[4], Joshua Robertson[1], Edward Wasige[2], Antonio Hurtado[1]

[1] Institute of Photonics, SUPA Dept of Physics, University of Strathclyde, Glasgow, United Kingdom
[2] High Frequency Electronics Group, University of Glasgow, Glasgow, United Kingdom
[3] International Iberian Nanotechnology Lab., Ultrafast Bio- and Nanophotonics Group, Braga, Portugal
[4] Centra-Ciências and Dept. de Física, Faculdade de Ciências, Universidade de Lisboa, Lisboa, Portugal

E-mail: antonio.hurtado@strath.ac.uk

## Abstract

We report a multi-modal spiking neuron that allows optical and electronic input and control, and wavelength-multiplexing operation, for use in novel high-speed neuromorphic sensing and computing functionalities. The photonic-electronic neuron is built with a micro-scale, nanostructure resonant tunnelling diode (RTD) with photodetection (PD) capability. Leveraging the advantageous intrinsic properties of this RTD-PD system, namely highly nonlinear characteristics, photo-sensitivity, light-induced I-V curve shift, and the ability to deliver excitable responses under electrical and optical inputs, we successfully achieve flexible neuromorphic spike activation and inhibition regimes through photonic-electrical control. We also demonstrate the ability of this RTD-PD spiking sensing-processing neuron to operate under the simultaneous arrival of multiple wavelength-multiplexed optical signals, due to its large photodetection spectral window (covering the 1310 and 1550 nm telecom wavelength bands). Our results highlight the potential of RTD photonic-electronic neurons to reproduce multiple key excitatory and inhibitory spiking regimes, at high speed (ns-rate spiking responses, with faster sub-ns regimes theoretically predicted) and low energy (requiring only ~10 mV and ~150 µW, electrical and optical input amplitudes, respectively), similar in nature to those commonly found in the biological neurons of the visual system and the brain. This work offers a highly promising approach for the realisation of high-speed, energy-efficient photonic-electronic spiking neurons and spiking neural networks, enabling multi-modal and multi-wavelength operation for sensing and information processing tasks, whilst also yielding enhanced system capacity, performance and parallelism. This work therefore paves the way for innovative high-speed, photonic-electronic, and spike-based neuromorphic sensing and computing systems and artificial intelligence hardware.

## 1. Introduction

Artificial spiking neurons and spiking neural networks (SNNs) are receiving increased research attention for applications in neuromorphic (brain-inspired) sensing and computing [1]. As a class of artificial neural networks (ANNs), which mimic some of the fundamental features of the brain, the key characteristics of SNNs include spiking neurons (nonlinear nodes) and synapses (connecting links) [2]. A biological spiking neuron can be abstracted using a simplified model which upon receiving inputs (stimuli) triggers a large output spike when the membrane potential reaches above a given critical threshold [3].

Several electronic devices have been identified as promising candidates for models that are capable of performing these (excitatory and/or inhibitory) neuronal characteristics observed in biology. These devices, such as those based on memristors [4-6], phase-transition processes [7,8] and magnetic tunnel junctions [9], propose exciting alternative electronic structures to those of conventional Von Neumann computing, yet still experience challenges arising from inherent obstacles such as the adverse impact of electromagnetic interference, and the fundamental trade-off between bandwidth and fan-in [10-12]. On the other hand, photonic approaches have emerged as an alternative to electronics-based systems. Photonic technologies offer a promising platform for the creation of high-speed, energy-efficient artificial spiking neurons and SNNs due to their inherent merits such as ultra-high bandwidth, low propagation losses, low crosstalk, high parallelism [13,14], and their high operation speeds (multiple orders of magnitude faster than the timescales in biological neurons [15]). Examples of recently reported photonic approaches to neuromorphic systems include a plethora of devices based on semiconductor lasers [16,17], optical modulators [18,19], and phase change material (PCMs) [20], to highlight a few (see review articles [21-23]). However, a key challenge in photonic approaches for artificial spiking neurons has been the implementation of inhibitory regimes controllably suppressing neuronal spike firing. Approaches requiring two parallel balanced photodetectors and optoelectronic conversion have been proposed [18-19], but simplified approaches reducing the number of components and complexity are required for the future development of neuro-inspired photonic spiking neural networks.

Apart from the aforementioned photonic approaches, research on emerging resonant tunnelling diodes (RTDs) has been rapidly developing [24-34]. By harnessing the highly nonlinear I–V curve of RTDs, a rich variety of neuromorphic behaviours, including excitability, spiking, multi-pulse bursting, and chaos [24-26], have been reported on isolated RTDs, or those with additional external photo-detectors (PDs) and coupled to laser diodes (LDs). Importantly, RTD-based systems with embedded photo-detecting capability (RTD-PD) allow intrinsic amplification due to their negative differential conductance (NDC) response [27-31]. Recent works have demonstrated that RTD neurons can deliver deterministic optically-triggered spiking, as well as two crucial neuromorphic properties for spike-based sensing and information processing: thresholding and refractoriness [28-31]. The well-defined spiking thresholding feature in RTD-PD neurons provides a reliable activation of all-or-nothing excitable electronic spiking responses, while the refractory period defines the upper limit of the spike firing rate [31]. However, recent studies have focused solely on the optically-induced neural-like spike firing in chip-scale RTD-PD neurons [31], without exploiting the versatile multi-modal photonic-electronic excitation capability of the system. This feature of RTD-PD neurons offers unique advantages, including the ability to elicit and dynamically control a wide range of additional neural-like behaviours, such as dynamical excitation and inhibition of fast spiking regimes and spike gating, crucial for their use as key elements in future optoelectronic neuromorphic sensing and computing platforms. Besides, RTD-PD neurons can absorb light signals at distinct wavelengths across a large spectral window, which for the systems used in this work cover the two key telecom wavelength windows centred at 1310 and 1550 nm in the infrared spectral region [31]. This inherent advantageous feature allows RTD-PD neurons not only to be applied to novel light-based neuromorphic sensing functionalities, but crucially also offer the ability to detect and integrate multiple, simultaneous wavelength-multiplexed optical inputs during operation. This is a key benefit of our photonic-electronic spiking neuron, when compared to other approaches for optical neuronal models, as it permits via wavelength-division multiplexing (WDM) protocols to increase the system's capacity through parallelised operation, and offer flexible routes towards the controlled achievement of light-triggered spiking regimes.

In this work, we demonstrate experimentally that RTD-PD spiking neurons can perform both spike activation and inhibition functionalities at high-speed rates by dynamically controlling the system operation using the multi-modal injection of photonic and electrical signals, as well as wavelength-multiplexed optical inputs. Crucially, all these neural-like spiking functionalities can be achieved using a single device removing the need for additional system components and providing a simple, versatile and compact platform for neuromorphic photonic (spike-based) sensing and processing platforms. Moreover, the chip-scale RTD-PD neurons of this work are designed with typical III-V compound semiconductor materials. The device consists of an electrical port for injecting direct current (DC) bias voltage and radio frequency (RF) signals, as well as a photoconductive spacer layer, making the device sensitive to optical illumination through its top window. This important feature permits the system to deliver neural-like spiking regimes upon the detection of optical signals within the a given spectral range, indistinctively of their wavelength. These unique features allow for RTD-PDs to operate as true photonic-electronic spiking sensory and computational neurons, able to elicit and inhibit a wide range of neural-like spiking regimes at high rates, and with low-power input signals, under both electrical and optical excitation (with simultaneous inputs at different wavelengths).

We fully capitalise on these key functionalities to demonstrate a wide range of photonic-electronic sensing-processing neuromorphic responses in RTD-PD spiking neurons. For instance, our results showcase that spike excitation can be achieved through electrical stimulus with the assistance of control optical signals. We also show that spike firing inhibition can be achieved by utilizing electrical signals as input stimuli and control optical signals for spike suppression. The same functionality can also be achieved by reversing the respective roles of these signals, using optical signals as stimuli for spike excitation and electrical input control signals for spike firing inhibition. We also analyse the scenario in which wavelength-multiplexed optical inputs arrive simultaneously in the RTD-PD spiking neuron. To that end, two distinct light beams at different wavelengths (1546 and 1552 nm are used in this work, at the telecom wavelength window centred around 1550 nm) are introduced into the RTD-PD neuron carrying each a different dynamical optical input. This enables the device to directly sense these optical signals and generate light-induced spiking regimes through optical wavelength multiplexing procedures. Moreover, by properly adjusting the polarity, optical power, and timing of the wavelength-multiplexed input signals, we demonstrate that this photonic-electronic RTD-PD spiking neuron permits complex sensing-processing functionalities. For example, we demonstrate that the first of the two applied wavelength-multiplexed optical channels can operate as the primary optical stimuli (for spike activation), while the second one serves as an auxiliary control signal to perform the excitation or inhibition, depending on the specific configuration of the input optical signals.

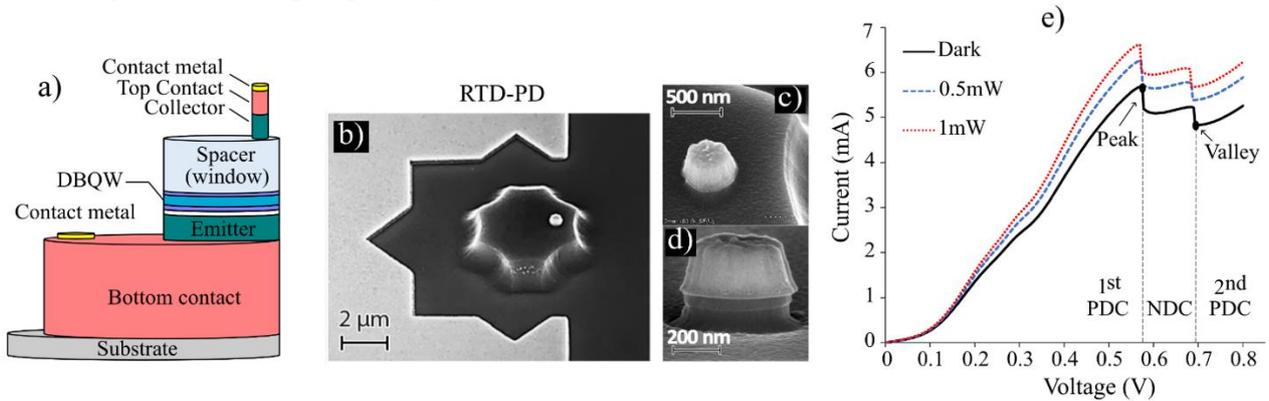

Fig.1 (a) Schematic layer structure (not to scale) of the RTD-PD device, featuring a nanostructure (nanopillar-shaped) top region. Light is injected into the photoconductive spacer layer vertically. (b)-(d) Scanning electron microscope (SEM) images of the RTD-PD, and the nanopillar-shaped structure mounted on top of the optical window (spacer) layer. (e) Current-voltage (I-V) curves of the RTD-PD measured under different conditions: in the dark (solid black line), and under 0.5 mW (blue dashed line) and 1 mW (red dotted line) of 1550 nm continuous-wave (CW) light. The "peak" and "valley" bias points are located at the boundaries of the Negative Differential Conductance (NDC) region, as indicated in the graph. The device is driven with reverse bias voltage.

## 2. Excitability in RTD-PDs photonic-electronic spiking neurons

This section describes the design of the RTD-PD and the mechanism that underpins the neuromorphic spiking sensing and processing responses achieved in these devices. The RTD-PD epitaxial wafer is developed on a semi-insulating InP substrate through metal organic chemical vapour deposition (MOCVD). The complete wafer layer composition can be found in [31]. Figs. 1(a) and 1(b-d) depict respectively the structure (Fig. 1(a)) and Scanning Electron Microscope (SEM) images of the RTD-PD of this work (Fig. 1(b-d)). At its core, the device contains a double barrier quantum well (DBQW), which consists of a 5.7 nm $In_{0.53}Ga_{0.47}As$ QW sandwiched by two 1.7 nm AlAs barriers. A 250 nm lightly doped InAlGaAs spacer layer was integrated on the upper side of the DBQW region to facilitate light absorption at infrared wavelengths. In order to decrease power consumption and promote energy-efficient spike generation, the layer structure above the upper spacer region is shaped into a nanopillar with a diameter of 500 nm [31].

RTDs exhibit unique characteristics that depend on their operating bias point, i.e. the applied bias voltage relative to its current-voltage (I-V) characteristic curve. Here, the RTD-PD was characterised using a reverse bias voltage. The I-V curve of the RTD-PD of this work, measured under dark conditions, is shown by the black solid line in Fig. 1(e). When biasing the device in either of the two positive differential conductance (PDC) regions in the I-V curve, the system exhibits a stable output response. However, when biasing the device in its negative differential conductance (NDC) region, the RTD circuit exhibits self-oscillations due to a presence of gain in the NDC. The latter behaviour in RTDs has been traditionally exploited for high-speed RF signal functionalities [32]. In turn, to operate the RTD as a neuromorphic photonic-electronic spiking neuron, the device is biased at either the near "peak" or near "valley" operation points, in close proximity to the boundaries of the NDC region in the I-V curve. For spike generation with an RTD driven by electrical input signals, RF signals can be applied in addition to the DC bias voltage [28-31] bringing the system over the threshold and into the NDC boundary in the I-V curve, triggering spiking events in response. The RTD-PD also embeds a 250nm-thick InAlGaAs spacer layer enabling photodetection [31]. Light absorption introduces carrier accumulation in regions adjacent to the DBQW region, leading to an observable optically-induced shift in the I-V curve [33-34]. The measured I-V curves of the RTD-PD when illuminated by Continuous Wave (CW) infrared light at 1550 nm, with an optical power of 0.5 mW (blue dashed line) and 1 mW (dotted red line), respectively, are shown in Fig. 1(e). As a result of the light-induced shift in the I-V curve, the operation point of the device can be changed in response to incoming optical signals, therefore enabling a controllable transition across the PDC/NDC boundary to deterministically trigger spiking regimes at the system's output granting a neuromorphic optical sensing functionality.

## 3. Multi-modal photonic-electronic RTD-PD neurons: hybrid activation and dynamical control of excitatory and inhibitory spiking regimes

The presence of two distinct and independent channels for the injection of electrical (RF) and optical stimuli allows RTD-PDs to offer unique features for neuromorphic spike sensing and processing. This section demonstrates experimentally three separate schemes, which make use of simultaneous multi-modal (photonic and electrical) stimulation to achieve dynamically controllable excitatory and inhibitory spike firing responses in the system.

The experimental setup utilised is illustrated graphically in Fig 2. CW light from a 1546 nm tuneable laser (TL) source is first passed through an optical isolator and a variable optical attenuator, respectively, to avoid undesired reflections and control the optical input power. After that, the TL's light is intensity-modulated by means of a Mach-Zehnder modulator (MZM) to encode the optical signal with input stimuli (RF signals) provided by an arbitrary waveform generator (AWG, channel 1). Prior to modulation, a polarisation controller (PC) is included in the setup to match the light polarisation of the TL's optical signal to that preferred by the MZM. The resulting modulated optical signal is injected into the optical window of the RTD-PD using a lens-ended optical fibre. For the electrical input stimulation line, an electrical (RF) signal generated also from the AWG (channel 2) is injected into the device (along with the DC bias voltage) using a bias-T. The dynamical response of the RTD-PD neuron (spike firing output responses) is collected via an RF power splitter and analysed with a real-time oscilloscope. To enable the RTD-PD neuron to exhibit spike firing regimes under external perturbations, the device is biased at the "peak" point in its I-V characteristic, close to the onset of the NDC region, with a reverse bias voltage of 577 mV. It should be noted that the closer to the boundary of the NDC region the applied bias voltage is, the lower is the required strength of the optical or electrical input signals (stimuli) used to trigger spike firing events in the system. Thus, adjusting the bias voltage represents a straightforward way to modify the threshold condition for spike firing excitation in the RTD-PD neuron. Importantly, the injected electrical or optical signals are only referred to as stimuli when they serve as perturbations that trigger spike firing events from the RTD-PD neuron. Otherwise, they are utilised as controls affecting the threshold condition (with minimal effect on spiking behaviour) to achieve the different sensing-processing functionalities presented in this work.

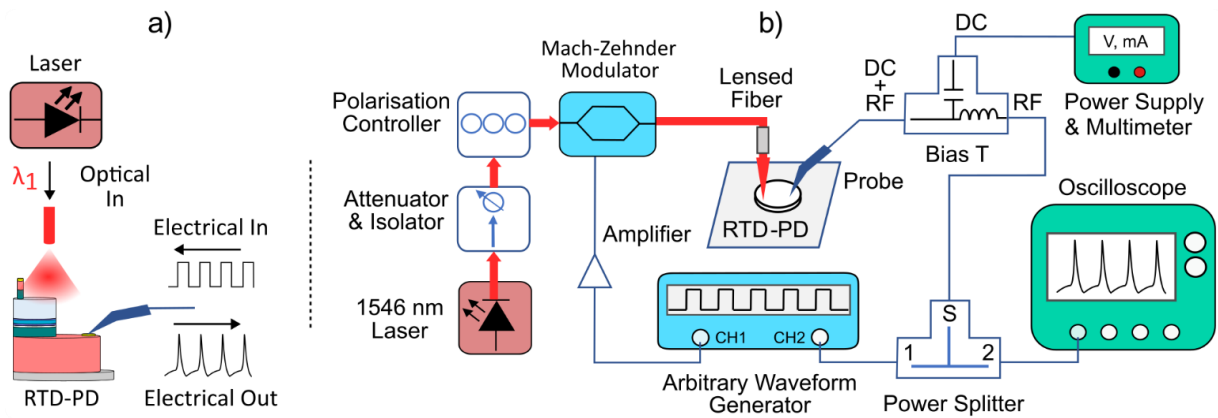

Fig 2. (a) Multi-modal operation with dynamically changing inputs in both the electrical and optical channel. (b) Experimental setup implemented to investigate the multi-modal photonic-electronic operation of an RTD-PD spiking neuron. The setup enables the injection of both optical (via a lens-ended fibre) and electrical (using a bias-tee) input signals for multi-modal operation. Either the optical or electrical input stimuli can be used to dynamically trigger or control excitatory and inhibitory spiking regimes in the RTD-PD neuron for use in neuromorphic sensing or processing functionalities.

In the first case, the RTD-PD spiking neuron is subject to electrical stimulation and optical control. Fig. 3 shows experimental results for this first scenario in which electrical input pulses (stimuli) are used to activate the firing of spike events in the RTD-PD neuron. Here the intensity modulated optical signal acts as control, tuning the RTD-PD neuron's spike firing threshold with light. The electrical input stimuli (see Fig. 3(a)) are sub-threshold (square-shaped) negative pulses with duration of 20 ns, drop amplitude of 10 mV, and a temporal separation between consecutive pulses of 500 ns. The optical input signal (see Fig. 3(b)) at 1546 nm controls the spiking threshold with a step-like waveform, with optical power that increases in step-like fashion from 0.04 to 0.16 mW. The extent of the light-induced shift in the I-V curve of the RTD-PD is determined by the intensity of the optical signal. This feature permits precise control over the activation threshold of the RTD-PD neuron via the intensity of the optically injected signal, controllably tuning the spike generation in the system. Fig. 3(c) depicts the electrical spiking output from the RTD-PD neuron. Here, the first three incoming sub-threshold electrical stimuli elicit no spiking responses as their intensity is below the system's activation threshold. However, as the intensity of the control/gating optical signal increases, there is a reduction in the spike firing threshold. This arises from the photocurrent generated by the optical signal which results in a reduction of the system's series resistance; hence shifting the peak point in the I-V curve towards lower voltages (see Fig. 1(c)). This enables the fourth input electrical stimuli, despite having the same amplitude as all previous electrical input pulses, to generate a spike event at the RTD-PD neuron's output. Fig. 3 therefore demonstrates the multi-modal photonic-electronic operation of the RTD-PD neuron, where the optical channel functions as a gate, providing additional capabilities for optically-induced controllable spike generation, sensing and encoding, which can be tuned by simply manipulating the optical intensity entering the system at any given time.

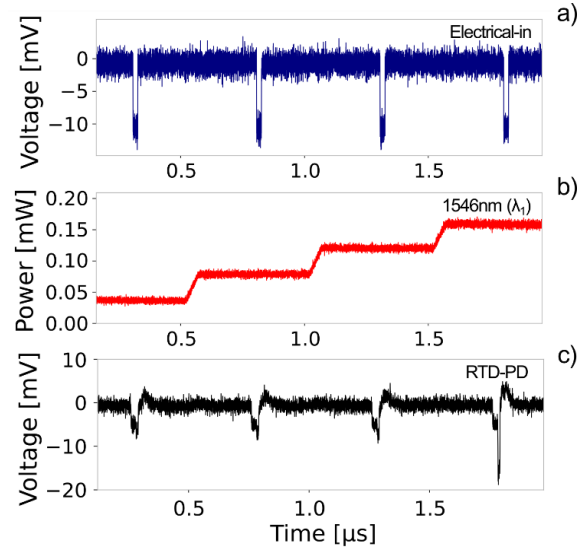

Fig. 3 Multi-modal operation of the RTD-PD neuron: Electrical stimulation with optical control. (a) Electrical input pulses act as stimuli for spike triggering and an (b) optical signal is used to adjust the activation threshold of the RTD-PD via the light-induced I-V shift effect. The optical input channel functions as a spiking switch or gating mechanism. (c) Output signal from the RTD-PD neuron showing spike activation when the electrical inputs simultaneously occur with the high-intensity optical gating signal.

The second case of analysis, shown in Fig. 4, demonstrates the operation of the RTD-PD neuron under the injection of dynamical input signals in both the electrical and optical channels towards controlled spike excitation and inhibition. Electrical input pulses (see Fig. 4(a)) with a duration of 20 ns and a separation of 500 ns are employed as stimuli to trigger spike firing in the system. The RTD-PD is (reverse) biased at the peak operation point with a voltage of 577 mV close to the NDC region boundary. In this situation, the RTD-PD neuron is expected to fire spikes in response to all electrical input stimuli. Optical input pulses (shown in Fig. 4(b)) with a power of 0.07 mW and a duration of 30 ns, matching the temporal locations of the electrical pulses, are also injected into the RTD-PD neuron. These optical pulses are set with an opposite polarity (to that of the electrical stimuli) to shift the current-voltage (I-V) curve in the opposite direction, away from the activation threshold of the system. Figure 4(c) shows the output signal from the RTD-PD, revealing that the system deterministically fires spike events in response to the injection of the electrical input stimuli only. This can be seen in the responses obtained for the second and third input electrical stimuli. In the case of the remaining inputs, the simultaneous arrival of an optical pulse suppresses the spike generation that should be elicited at the output of the RTD-PD neuron. Instead a small amplitude change, due to the optical photo-response and input current leakage, is obtained instead, as can be seen in Fig. 4(c). To achieve spike inhibition, it is necessary to provide sufficient optical power to counteract the I-V shift induced by the electrical input. It is important to note that the amplitudes of the electrical stimuli and optical input are also dependent on the operating point (bias voltage) of the device, as it directly defines the spike firing threshold. Importantly, Fig. 4 not only demonstrates the high-speed dynamical operation of the RTD-PD neuron for sensing and processing operations and under both optical and electrical inputs, but the multi-modal controllable excitation and inhibition of spiking regimes in the system.

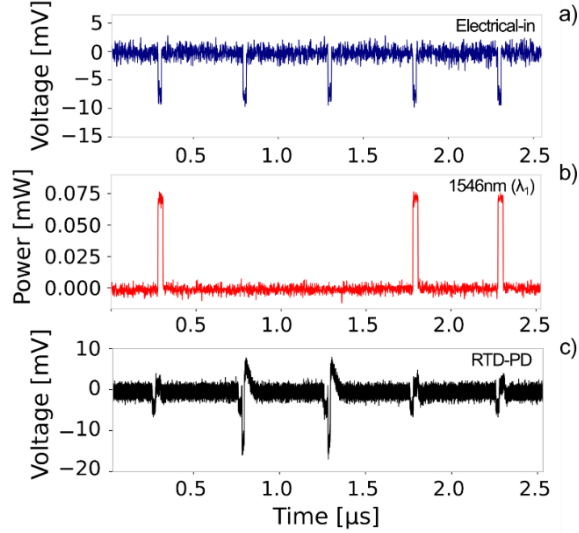

Fig. 4. Multi-modal dynamical excitation and inhibition of spiking regimes in an RTD-PD neuron: Electrical stimulation with optical control. (a) Electrical input pulses of negative polarity act as stimuli for spike triggering in the system. (b) Optical input pulses with positive polarity (opposite to that of the electrical stimuli), with large enough optical power (to counteract the effect of the electrical stimuli) are simultaneously injected into the device. (c) The output signal from the RTD-PD showing controlled inhibition of individually-addressed spike firing events.

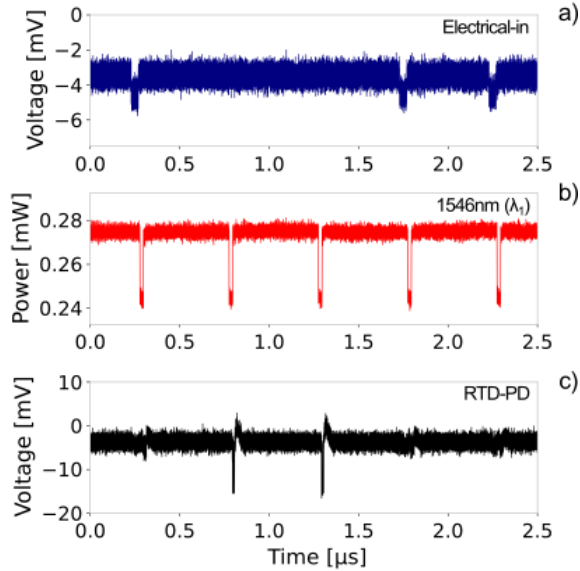

Fig. 5. Multi-modal dynamical excitation and inhibition of spiking regimes in an RTD-PD neuron: Optical stimulation with electrical control. (a) Sub-threshold electrical input pulses with negative polarity are injected into the system. (b) Optical input pulses act as stimuli for spike triggering in the system. (c) The simultaneous injection of the optical and electrical pulses counteracts the temporal light-induced I-V curve shift and produce controlled inhibition of individually-addressed spike firing events at the output of the RTD-PD.

Figure 5 further highlights the multi-modal, versatile, sensing-processing operation of the photonic-electronic RTD-PD neuron. Fig. 5 shows results for a different scenario in which optical input pulses are used as excitatory (sensory) stimuli to trigger spike firing events in the system, whilst electrical inputs dynamically inhibit the optically-activated spiking responses. The RTD-PD is biased at the peak with a (reverse bias) voltage of 575 mV and input optical pulses (stimuli) with negative polarity (shown in Fig. 5(b)) are used to deterministically trigger spikes in the system [31]. Hence, in Fig. 5(c), the output from the RTD-PD neuron fires spiking responses when only optical stimuli are applied (see the responses to the 2nd and 3rd optical input pulses). In parallel, electrical pulses with negative polarity (shown in Fig. 5(a)) and low sub-threshold amplitude (insufficient to trigger spiking responses themselves in the system) are synchronously injected. These enable direct processing tasks to be performed in the system by counteracting the temporal shift in the I-V curve produced by the optical input pulses, thereby inhibiting the optically-induced spike firing events. Owing to the inhibitory effect of the electrical input, the 1st, 4th, and 5th optical pulses are unable to trigger spike events in the system, and small non-excitable outputs are observed instead.

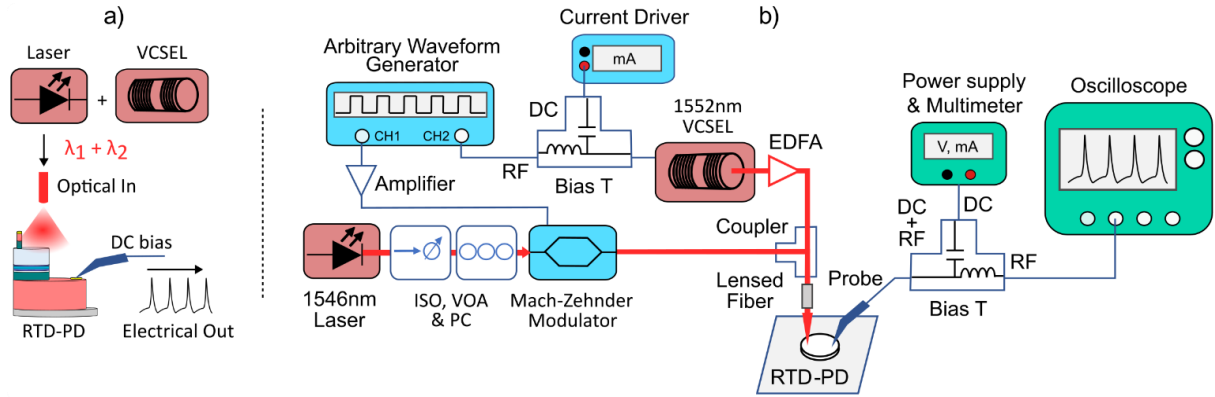

Fig. 6 (a) Multi-wavelength operation of the RTD-PD sensing-processing spiking neuron with dynamical inputs from two optical channels. (b) Experimental set-up: two optical signals are generated from a TL and a VCSEL at 1546 and 1552 nm, respectively. Each independent photonic input (sensory) signal is encoded with optical stimuli and wavelength-multiplexed into a single optical channel for their combined optical injection into the RTD-PD neuron. Different light-triggered spike sensing-processing operations are demonstrated under multi-wavelength optical injection, including spike activation and inhibition regimes.

## 4. Multi-wavelength operation of RTD-PD photonic-electronic spiking neurons

A key aspect of the RTD-PD neuron of this work is its photo-detection capability with sensitivity across a wide region in the infrared spectrum (covering at least the two telecom wavelength windows centred at 1310 and 1550 nm). This feature not only opens its potential use in novel light- and spike-based neuromorphic sensing functionalities, but also offers a crucial advantage, i.e. multiple wavelength-multiplexed optical signals can be injected in parallel, being all simultaneously absorbed by the RTD-PD neuron, eliciting spiking responses. In addition to the wavelength-insensitive and multi-wavelength operation capabilities of the system, it is also possible to manipulate the power and polarity of the incoming optical stimuli to provide additional optical-induced spike sensing and processing functionalities. Figure 6 depicts the experimental setup used to investigate the spiking regimes in the RTD-PD neuron under multi-wavelength optical operation. Specifically, two independent optical signals were used at the wavelengths of 1546 and 1552 nm, right at the centre of the so-called C-Band of the optical telecommunications spectrum. These were generated with a Tuneable Laser (TL) and a Vertical Cavity Surface Emitting Laser (VCSEL), respectively. Optical input pulses were encoded in each channel (at 1546 and 1552 nm, respectively) by externally modulating the TL's light and by directly modulating the bias current of the VCSEL (using a Bias-T). The optical output of the modulated VCSEL was then amplified using an Erbium-Doped Fibre Amplifier (EDFA) to provide additional optical power. After this, the two optical input channels were wavelength-multiplexed into a single optical fibre using a fibre coupler before optical injection through a lens-ended optical fibre into the RTD-PD neuron. The latter was biased at the peak operation point with a reverse bias voltage of V = 581 mV.

Figure 7 shows results demonstrating the multi-wavelength optical spike triggering operation of the RTD-PD sensing-processing neuron of this work. Two optical input signals at 1546 nm (channel 1) and at 1552 nm (channel 2) with encoded super-threshold optical stimuli are generated. These are shown respectively in the time-series plots in Figs. 7(a) and 7(b), in which the optical input stimuli are marked with shaded boxes. Also depicted in Figs. 7(a) and 7(b) are the measured output signals from the RTD-PD spiking neuron, in response to each individual optical input at each wavelength. Since all encoded optical stimuli have super-threshold intensities, the RTD-PD neuron fires a spike event after detecting (sensing) the arrival of every optical stimulus in each independent channel. Figure 7(c) shows in turn the case in which both optical input channels are wavelength-multiplexed into a single optical line and injected simultaneously into the device. These results show that the RTD-PD indeed operates under wavelength-multiplexed inputs, eliciting sequentially spiking events at its output upon detecting the optical stimuli in both parallel inputs at 1546 nm (greed shaded) and 1552 nm (blue shaded). The latter shows that all-or-nothing electrical spikes are produced with the same amplitude for each optical channel, despite the differing wavelengths.

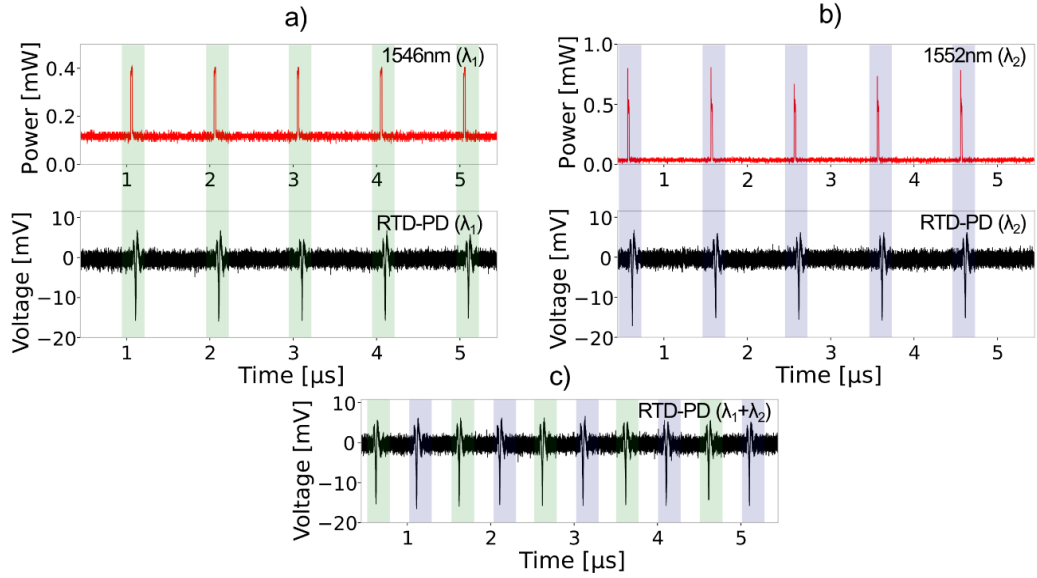

Fig.7. Multi-wavelength optically-triggered deterministic spike firing in an RTD-PD sensing-processing neuron. Optical input signals at (a) 1546 and (b) 1552 nm are both encoded with super-threshold optical stimuli, shaded in green (a) and blue (b) respectively. (a) and (b) have corresponding time-series that show the output of the RTD-PD neuron when each is injected individually into the device. (c) Time-series measured at the output of the RTD-PD neuron when both optical input channels at 1546 nm and 1552 nm are wavelength-multiplexed and injected simultaneously into the device. The generated spiking pattern demonstrates the successful operation of the RTD-PD neuron under the arrival of wavelength-multiplexed optical stimuli, with spikes generated by both the 1546 nm and 1552 nm input optical signals.

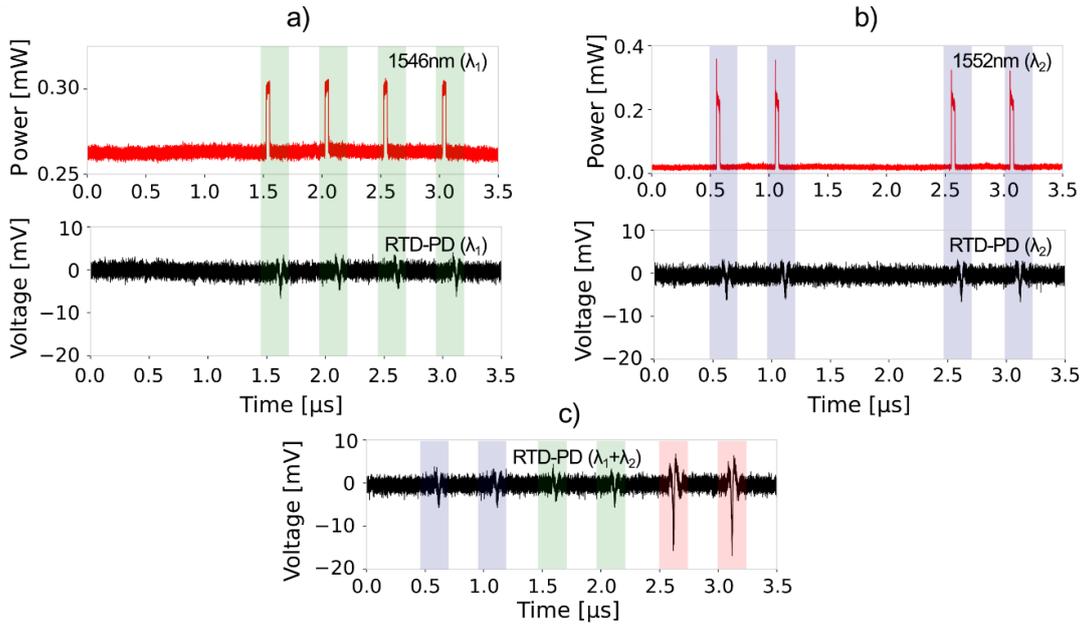

Fig. 8 Multi-wavelength optically-induced spike activation in the RTD-PD sensing-processing neuron through optical stimuli enhancement. Optical input signals at the wavelengths of 1546 nm (a) and at 1552 nm (b), both encoded with sub-threshold optical stimuli at different time instants (shaded in green and blue). Each optical input (a) and (b) show the corresponding measured time-series at the output of the RTD-PD neuron when each signal is injected individually into the device. For both cases no spikes are fired for the sub-threshold optical inputs. (c) Time-series measured at the output of the RTD-PD neuron when both optical input channels at 1546 and 1552 nm are wavelength-multiplexed and optically-injected simultaneously into the device. The system fires optical spikes (red shaded boxes) only when the (sub-threshold) stimuli at both wavelengths arrive coincidentally and their combined energy pushes the system over the spiking threshold. The generated spiking pattern demonstrates successful integration by the RTD-PD sensing-processing neuron of multi-wavelength sub-threshold optical stimuli.

The RTD-PD neuron's ability to sense and process multiple optical wavelength-multiplexed signals also enables the generation of spike firing events through multi-wavelength optical injection. Here, a boosting effect can be produced by multiple coincidental wavelength-multiplexed, sub-threshold optical stimuli. Where combined energies exceed the system's activation threshold, even individual optical inputs (encoded at different wavelengths) with low insufficient amplitudes, can lead to the firing of excitable spiking responses. Figure 8 demonstrates the multi-wavelength controllable firing capability of the RTD-PD neuron. Depicted in Figs. 8(a) and 8(b) respectively, are two configured optical inputs at the wavelengths of 1546 nm and 1552 nm, with sub-threshold optical stimuli at different time-instants. Figs. 8(a) and 8(b) also show respectively the temporal responses at the output of the RTD-PD neuron when the two optical input signals are injected individually into the device. Since all encoded stimuli in both optical channels have amplitudes below the threshold required to trigger a spiking response in the system, the output of the RTD-PD is non-spiking and akin to a photodetector generating only small amplitude photo-responses upon the arrival of optical stimuli. By wavelength-multiplexing the two optical channels and injecting them simultaneously into the device, a completely different response is elicited by the RTD-PD sensing-processing system, as demonstrated in Fig. 8(c). Now, when the two sub-threshold optical stimuli at the two different wavelengths overlap in time, their combined energies are integrated in the device due to its photo-sensing capability. As a result, the total integrated energy of the coincidental sub-threshold optical stimuli exceeds the excitable activation threshold of the RTD-PD neuron, which fires as a result the spike events marked in red shaded boxes in Fig. 8(c). The latter also shows that as expected when a single sub-threshold optical stimulus enters the RTD-PD neuron, only a small amplitude (non-spiking) photo-response is obtained. Therefore, Fig. 8(c) clearly demonstrates the system's ability to detect, integrate and process multiple wavelength-encoded independent light stimuli, readily allowing the RTD-PD neuron to deliver sensing-processing tasks with multiple wavelength-multiplexed optical inputs. Such functionality is essential for enabling increased system parallelism, large fan-in and complexity. It must be also noted that whilst the demonstration in this work is provided with two optical input signals with similar optical power at two close wavelengths in the 1550 nm region, the system is completely flexible permitting light absorption in a large infrared spectral window (covering at least the two telecom wavelengths bands at 1310 and 1550 nm).

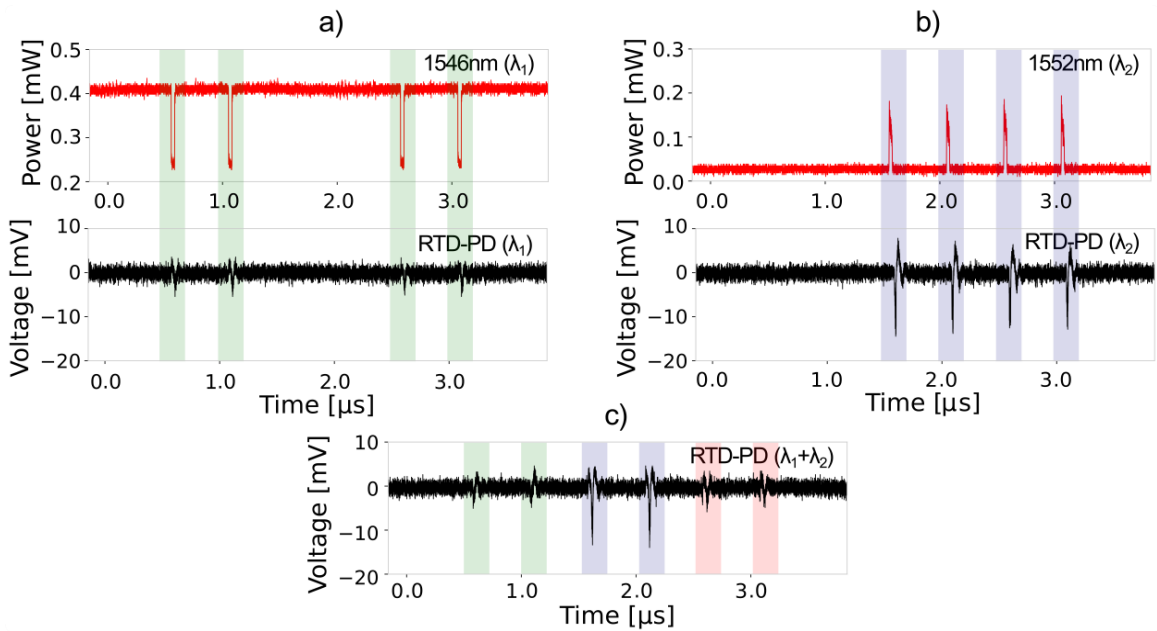

Fig. 9. Multi-wavelength optical spike inhibitory dynamical regimes in the RTD-PD sensing-processing neuron. Optical input signals at the wavelengths of (a) 1546 nm and (b) 1552 nm, encoded respectively with sub-threshold optical pulses with negative polarity, and super-threshold optical pulses with positive polarity. Each input (a) and (b) show the corresponding temporal response at the output of the RTD-PD when each signal is injected independently. As expected, the sub-threshold stimuli at 1546 nm only cause small amplitude photo-responses in the system (shaded in green), whilst the super-threshold optical stimuli at 1552 nm elicit the firing of excitable spikes (shaded in blue). (c) Measured temporal response at the RTD-PD neuron when both optical signals are wavelength-multiplexed and injected simultaneously into the device. The response demonstrates spike inhibition when a negative polarity sub-threshold optical pulse arrives coincidentally with a super-threshold optical stimulus (shaded red boxes). The non-coincident pulses trigger the expected responses at the output of the RTD-PD.

Just as the multi-modal photonic-electronic stimulation of the RTD-PD neuron can be used to produce spike inhibitory effects (see Figs. 4 & 5), the use of multiple wavelength-multiplexed optical inputs also permits the achievement of inhibitory sensing-processing dynamical functions. This requires now the simultaneous injection of appropriate optical stimuli (at different wavelengths) with opposite polarities. Figure 9 demonstrates experimentally the multi-wavelength spike inhibitory dynamical operation of the RTD-PD neuron. The first optical input signal at 1546 nm (shown in Fig. 9(a)) is encoded with sub-threshold, negative polarity, pulses. Figure 9(a) shows that these optical inputs produce no spike firing outputs at the RTD-PD neuron when injected individually into the device. As expected, the sub-threshold optical stimuli produce a small-amplitude photo-response, as shown in Fig. 9 (b). The second optical input signal at 1552 nm is shown in Fig. 9(b). This optical input is encoded with super-threshold optical stimuli of positive polarity that in turn produce a spiking response at the RTD-PD neuron's output when injected independently into the device. Here, as expected, the optically-injected stimuli had sufficient energy to cross the activation threshold of the RTD-PD neuron. The two optical input signals were then wavelength multiplexed, combined in a single optical fibre line and fed into the device simultaneously, producing the RTD-PD response shown in Fig. 9(c). This shows that when single pulses from either of the two optical signals arrive individually, the same responses are observed: negligible amplitude non-spiking photo-responses for the arrival of the negative polarity sub-threshold optical pulses at 1552 nm, and excitable spikes for the arrival of positive polarity super-threshold optical stimuli at 1546 nm. However, Fig. 9(c) experimentally demonstrates the achievement of a novel multi-wavelength optically-induced spike inhibitory response when the optical stimuli from both signals at 1546 and 1552 nm arrive coincidentally into the RTD-PD neuron. In this last case (red shaded), when both optical stimuli overlap in time, due to their opposite polarities, they counteract each other and the total optical power entering the RTD-PD is lowered. This reduces the magnitude of the temporal (optically-induced) shift in the device's I-V curve, to the point that this is not sufficient to exceed the system's spike firing excitation threshold. As a result, the combined effect of both opposite optical input stimuli is analogous to that produced by a single sub-threshold optical perturbation, and the RTD-PD therefore remains in a quiescent state. Consequently, due to the counterbalancing effect of the combined optical input pulses with opposite polarities, the system yields a controllable spike inhibitory response. These findings validate the capability of RTD-PD neurons to operate as excitatory, as well as inhibitory, spiking sensing-processing nodes thanks to its photon-enabled and wavelength-multiplexed operation Optical wavelength multiplexing therefore serves as an efficient method of fully utilizing the capacity of the optical system, with enhanced flexibility to control the behaviour of RTD-PD photonic neurons for sensing and processing tasks without the need for extra power-consuming optoelectronic components.

## 5. Conclusion

This work reports experimentally on a multi-modal, multi-wavelength, photonic-electronic spiking neuron enabling neuromorphic sensing and processing functionalities, and based upon a micro-scale, nanostructure RTD-PD. We demonstrate experimentally multiple functions for the triggering of excitatory and inhibitory spiking regimes in these key-enabling RTD-PD photonic-electronic neurons through hybrid optical and electrical control, and wavelength-multiplexed optical stimuli. The electrical control of the RTD-PD neuron equips the device with the ability to tune the elicitation/suppression of spike firing regimes by dynamically modulating the system's excitability threshold. Meanwhile, the RTD-PD is comprised of a photoconductive spacer layer with an embedded optical window enabling sensitive photodetection across a large infrared spectral range, yielding light-driven shifts in the device's I-V characteristic that can stimulate the detection (sensing) of optical signals and deliver the firing of optically-triggered spikes. In this work, we focus on the multi-modal photonic-electronic hybrid control of this RTD-PD sensing-processing neuron. We demonstrated that the injection of optical input signals can function as an effective spike gating switch, which effectively augments the effect of sub-threshold electrical stimuli to deterministically triggered spike events in the system. This is achieved by the temporal light-induced shifts in the I-V curve caused by the optical inputs, which reduce the threshold for spike activation allowing otherwise sub-threshold electrical stimuli to elicit spike firing events. Additionally, we also show that spike inhibition dynamical responses can also be triggered in the system under the injection of dynamical multi-modal photonic-electronic inputs. For spike inhibition, we show that both optical or electrical control inputs can be respectively used to supress spike firing regimes caused by super-threshold electrical and optical stimuli. Specifically, to accomplish effective spike inhibition, the inhibitory signal should possess enough amplitude to produce an opposing I-V curve shift that counteracts the transition induced by the (optical or electrical) super-threshold stimuli. Moreover, both the activation and inhibitory inputs should overlap in time to permit the controllable, dynamical, suppression of individually-targeted ns-rate spikes in the system. We also demonstrate the ability of the RTD-PD spiking neuron to simultaneously detect and operate with multiple independent wavelength-multiplexing optical input signals encoded at different wavelengths (1546 and 1552 nm in the telecom C-band are used in this work). We demonstrate multiple versatile sensing-processing functions enabled by the unique multi-wavelength capability of the system. This includes first the demonstration of wavelength-multiplexed operation to deliver parallel optical excitation and increased system capacity. We also demonstrate the system's ability to integrate multiple sub-threshold optical inputs encoded in different wavelengths. Here all-or-nothing spike events were triggered only when the total combined input intensities exceed the activation threshold of the RTD-PD. Finally, we showed the capability to inhibit spike firing regimes by using multi-wavelength optical input pulses of opposite polarities. The unique capability of optical wavelength multiplexing for spike generation and dynamical control, makes full use of the high bandwidth and parallelism present in optical communication networks, and highlights the potential for full optical control of RTD-PD neurons devices forming neuromorphic photonic-electronic sensing and computing modules and SNNs. Importantly, all the spike (excitatory and inhibitory) functions demonstrated experimentally, are obtained using high speed (up to ns-rates in this work, but sub-ns operation should be possible in future optimised systems [28]), and low-energy optical and electronic inputs (down to ~10s mVs and sub-mW electrical and optical input signals). These results offer excellent promise for future compact (micro-/nano-scale), fast (multi-GHz), energy efficient ($10^{-13}$J/spike) [28] hybrid photonic-electronic, wavelength-multiplexed RTD-based spiking neurons and SNNs for future light-enabled neuromorphic sensing and computing technologies and artificial intelligence hardware.

## Acknowledgements

The authors acknowledge support by the European Commission (Grant 828841-ChipAI-H2020-FETOPEN-2018-2020) and by the UK Research and Innovation (UKRI) Turing AI Acceleration Fellowships Programme (EP/V025198/1). The authors would also like to acknowledge IQE plc. for providing the semiconductor wafers used to fabricate the systems of this work.